\documentclass{aip-cp}
\usepackage[numbers]{natbib}
\usepackage{rotating}
\usepackage{graphicx}

\def\ket#1{  \left\vert  #1   \right\rangle   }

\begin{document}

\title{Structural Complexity Of Quantum Networks}

\author[aff1,aff2]{Michael Siomau}

\eaddress{siomau@nld.ds.mpg.de}

\affil[aff1]{Physics Department, Jazan University, P.O.Box 114,
45142 Jazan, Kingdom of Saudi Arabia}

\affil[aff2]{Network Dynamics, Max Planck Institute for Dynamics and
Self-Organization (MPIDS), 37077 G\"{o}ttingen, Germany}

\maketitle

\begin{abstract}
Quantum network is a set of nodes connected with channels, through
which the nodes communicate photons and classical information.
Classical structural complexity of a quantum network may be defined
through its physical structure, i.e. mutual position of nodes and
channels connecting them. We show here that the classical structural
complexity of a quantum network does not restrict the structural
complexity of entanglement graphs, which may be created in the
quantum network with local operations and classical communication.
We show, in particular, that 1D quantum network can simulate both
simple entanglement graphs such as lattices and random graphs and
complex small-world graphs.
\end{abstract}

\section{INTRODUCTION}

Modern world exhibits complex structures, which can be hardly
tackled with simple linear models. Nonlinear networks
\cite{Dorog:03}, in contrast, can reliably describe a large portion
of the structures and their dynamics. So far, networks have been
successfully applied to describe such different natural and
artificial processes as epidemic spreading \cite{Capasso:93},
dynamics and self organization of neural circuits \cite{Finger:01},
intra-social interactions \cite{Easly:10} and even autonomous
control of robots \cite{Steingrube:10} to name just a few. Networks
permeate various structures, where different parties interact with
each other. Differences in the characteristics of such interactions
and how they evolve in time give growth to different types of
structures: \textit{regular} and \textit{random} networks or,
filling the gap between aforementioned, \textit{complex} networks,
which do not have a regular structure but neither are completely
random \cite{Boccalettia:06}.

While it is typically implied that networks obey the laws of
classical physics, quantum networks have come into focus recently
due to remarkable advances in quantum technologies. For instance,
the entanglement-based quantum communication has been performed for
distances over 144 km \cite{Ursin:07}, which makes feasible the long
distance quantum communication \cite{Duan:01} and the quantum
internet \cite{Kimble:08}. Also, latter achievements in experimental
implementations of quantum computers \cite{Ladd:10} implies that
quantum devices with limited computational capabilities could become
available in the nearest future, which opens the possibility to
realize the distributed quantum computation \cite{Cirac:99}. These
emerging quantum systems are of non-trivial structure and need to be
treated from the network perspective.

A study of a network, be it classical or quantum, always begins with
defining its structure. Following classical paradigmatic viewpoint,
structural complexity of quantum networks is typically associated
with its physical structure, i.e. spatial position of nodes and
their inter-connectivity due to channels. But, quantum mechanics as
a non-local theory offers an intriguing possibility to prepare
quantum systems located at physically disconnected nodes into an
entangled state. This preparation can be accomplished with local
operations and classical communication (LOCC) \cite{Nielsen:00}
without direct interaction of the distant nodes. This fact alone
makes the difference between the physical structure of the quantum
network (i.e. nodes and channels) and it's entanglement structure
(i.e. nodes and entanglement links). In this paper we aim to make
the difference explicit.

To show the distinction between classical and quantum structural
complexity of quantum networks, we shall consider the simplest
possible 1D network configuration, where nodes are placed on a line
at fixed distance from each other and are connected by channels. We
assume that the nodes possess arbitrary number of qubits together
with measurement devices and communicate with each other through the
channels. For the sake of simplicity, we assume that only two-qubit
entangled states can be created connecting two different neighboring
nodes and no decoherence effects are taking place. We employ a
particular type of LOCC -- the entanglement swapping \cite{Pan:98}
-- to show that non-trivial entanglement graphs (\textit{egraphs})
can be created on the 1D quantum network. In the next section, we
introduce basic notations and the entanglement swapping operation.
Then we show what kind of egraphs can be simulated on the 1D network
and at what cost in terms of initial resources. We conclude in the
last section putting forward open questions.

\section{ENTANGLEMENT SWAPPING}

Exchanging photons and classical information, two qubits at
neighboring nodes can be prepared in a two-qubit entangled state,
which form an entanglement link between the nodes. Let us assume
that the state is pure and thus may be written in the computational
basis \cite{Nielsen:00} as
\begin{equation}
\label{states}
 \ket{\varphi} = \sqrt{\lambda_1} \ket{00} + \sqrt{\lambda_2} \ket{11}
 \, ,
\end{equation}
where $\lambda_1$ and $\lambda_2$ are the Schmidt coefficients
conditioned by $\lambda_1 \geq \lambda_2$ and $\lambda_1 + \lambda_2
= 1$.

If a node hosts two qubits from two different entangled pairs as
depicted in Fig.~1a on the top, the qubits at the node may be
measured in the Bell basis
\begin{eqnarray}
 \label{Bell-measurements}
\ket{\Psi^{\pm}} = \frac{1}{\sqrt{2}} \left( \ket{00} \pm \ket{11}
\right) \, , \hspace{1cm} \ket{\Phi^{\pm}} = \frac{1}{\sqrt{2}}
\left( \ket{01} \pm \ket{10} \right) \, ,
\end{eqnarray}
projecting the initial four-qubit state $\ket{\psi_{1234}} =
\ket{\varphi_{1,2}} \otimes \ket{\varphi_{3,4}}$ onto one of the
Bell states. This leads to "unification" of the two entanglement
links into a single link disconnecting the host node as shown in
Fig.~1a on the bottom.

In general, entanglement swapping reduces entanglement of the final
state in comparison to the initial entanglement of the two entangled
pairs \cite{Pers:08}. But, Bose, Vedral and Knight showed that there
is a measure of entanglement preserved {\it on average} under the
entanglement swapping \cite{Bose:99} for any pure state of qubits
(\ref{states}) -- the singlet conversion probability defined as $p =
2 \lambda_2$. This remarkable result allows us to assume that an
arbitrary number of entangled links can be united into a single link
by entanglement swapping with no detrimental effect to the
entanglement of the final link. This assumption is crucial for
further discussion and allows us to consider all the entanglement
links in the egraph as equivalent, irrespectively whether they are
created between neighboring (due to a photon exchange) or distant
(due to entanglement swapping) nodes.

\section{QUANTUM EGRAPHS IN 1D QUANTUM NETWORK}

It is evident from the Fig.~1a that the entanglement swapping
operation "consume" two local entanglement links to create a single
non-local entanglement link. Thus, construction of any non-trivial
egraph on the 1D quantum network requires defined initial resources
-- a number of local entanglement links (which are created by
sending a photon between two neighboring nodes). Let us assume that
$k$ entangled pairs are initially created between each pair of
neighboring nodes. Let us find out how the total number of the
initial entangled pairs in the network $K = k N$ depends on the
complexity of simulated egraphs.

A ring can be simulated on the 1D network at cost of just $K=2N$
local entanglement links by applying entanglement swapping once at
each node except of the borders (see Fig.~1b).

A 2D squared lattice of $n \times n = N^2$ nodes (see Fig.~2) can be
simulated on the 1D network at the cost of
\begin{eqnarray}
 \label{2D cost }
K_{2D} = (n-1) \left[ \sum_{i=1}^{n-1} (2^i + 1) \right] + n^2 -1  =
(n-1) \; 2^n + 2(n-1)^2  =  (\sqrt{N}-1) \; 2^{\sqrt{N}} +
2(\sqrt{N}-1)^2
\end{eqnarray}
initial entangled pairs. The number of initial entangled pairs
growth exponentially with the lattice size, which makes the
simulation impractical.

This situation improves radically if we try to simulate a random
graph with the 1D network. An arbitrary Erdos-Renyi random graph can
be constructed from the complete graph \cite{Capasso:93}, if the
links are present/removed with some probability. If we can simulate
the complete egraph on the 1D, i.e. the graph where every pair of
nodes is connected by an entanglement link, then we can also
simulate an arbitrary random egraph by destroying entanglement
links, for example, by measuring one of the qubits from the
entangled pair forming the entanglement link in the computational
basis. The complete graph of $N$ nodes (see Fig.3) can be created
with
\begin{eqnarray}
 \label{2D cost }
K_{RG} =  \sum_{i=1}^{N-1} i (N-i) = \frac{(N-1)N(N+1)}{6}
\end{eqnarray}
initial entangled pairs. The number of initial entangled pairs
growth polynomially with the number of nodes. This makes practical
simulations feasible. Interestingly, because the complete graph of
$N$ nodes has exactly $N(N-1)/2$ links, on average $(N+1)/3$ local
entanglement links are needed to construct an entanglement link of
the complete egraph.

Finally, we would like to give an example of a small-world complex
graph \cite{Watts:98}, which may be simulated on the 1D quantum
network. Fig.~4 shows a hierarchical small-world egraph (which is
also a fractal due to self-similarity) \cite{Boettcher:09}. The
construction of the egraph requires surprisingly small initial
resources $K_{H-SW} = N (1 + \log_2 (N-1))$, which makes it very
useful in quantum communication due to its remarkable percolation
properties \cite{Siomau:16}.

\begin{figure}
 \includegraphics[width=1\textwidth]{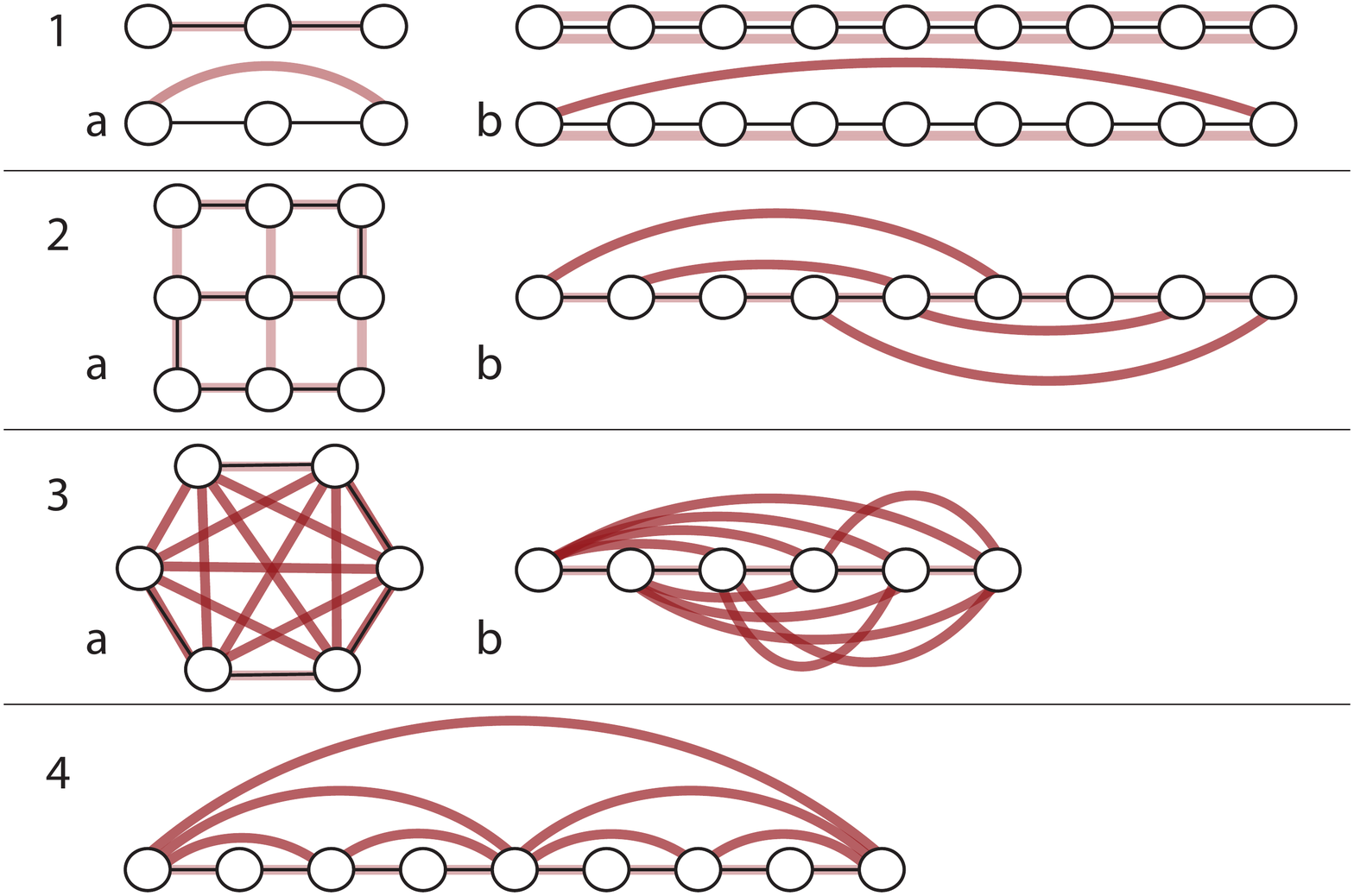}
\caption{Structural complexity of 1D quantum network: black lines
show the channels, red lines show entanglement links. 1a:
Entanglement swapping at the central node connects the neighboring
nodes with a non-local entanglement link. 1b: A ring is created from
the 1D network by multiple entanglement swapping. 2: A 2D squared
lattice of entanglement links: (2a) folded and (2b) planar
representations of the egraph. 3: A complete egraph: (3a) folded and
(3b) planar representations. 4: A small-world egraph.}
\end{figure}

\section{CONCLUSION}

We showed a new feature of quantum networks -- the inequivalence
between classical and quantum structural complexity of the network.
Already the 1D quantum network is able to simulate lattices and
random and small-world graphs. The simulation of the 2D egraph
requires exponential initial resources, thus is inefficient. In
contrast, polynomial growth of the initial resources with respect to
the number of nodes manifests that random and small-world egraphs
can be simulated efficiently with 1D quantum network. This opens
intriguing possibilities to study non-trivial egraphs and their
dynamics with just a linear 1D setup.

Although our results are concerned with undirected egraphs, they may
be extended to directed and weighted \cite{Boccalettia:06} egraphs.
The direction may be introduced in the egraph model due to classical
communication between the nodes by restricting some nodes from
sending classical information to others. Such nodes become in-nodes,
while those able to send classical information are out-nodes. The
normalized weights of the entanglement links are essentially
represented by the amount of entanglement preserved in the link. The
singlet conversion probability of an entanglement link can be
reduced by local operations or decoherence and increased by
entanglement distillation \cite{Nielsen:00}.

General questions arise from our results: Can an arbitrary egraph be
simulated on the 1D quantum network by LOCC? Under what conditions
this simulation is efficient in terms of initial resources? Further
studies are needed to answer the questions.

\section{ACKNOWLEDGMENTS}
This work was supported by KACST under research grant no.~180-35.

\end{document}